\documentclass[conference]{IEEEtran}

\usepackage{multirow}
\usepackage{amsfonts}
\usepackage{amssymb}
\usepackage{latexsym}
\usepackage{cite}
\usepackage[tight,footnotesize]{subfigure}
\usepackage{color}
\usepackage{mathtools}
\usepackage{amsmath}
\usepackage{dblfloatfix}

\usepackage[utf8]{inputenc}
\usepackage[autostyle]{csquotes}
\usepackage{textcomp}
\DeclareMathOperator{\E}{\mathbb{E}}
\usepackage[font=small,belowskip=-10pt]{caption}

\usepackage{siunitx}

\def\footnoterule{\relax
	\kern-5pt
	\hbox to \columnwidth{\hfill\vrule width 0.5\columnwidth height 0.4pt\hfill}
	\kern4.6pt}

\newcommand{\etal}{\textit{et al.}}
\newcommand{\apo}{\textquotesingle}

\newcommand{\xppp}{\phi_X^P}
\newcommand{\yppp}{\phi_Y^P}
\newcommand{\pppp}{\phi_p^P}
\newcommand{\vppp}{\phi_v^P}
\newcommand{\nlos}{n_{L}}
\newcommand{\nnlos}{n_{NL}}
\newcommand{\pcp}{\phi^C_v}

\newcommand{\texp}{\text{exp}}
\newcommand{\lr}{\lambda_r}
\newcommand{\lv}{\lambda_v}
\newcommand{\lpar}{\lambda_p}
\newcommand{\cbar}{\overline{c}}
\newcommand{\csig}{\sigma_c}

\newcommand{\plf}{\ell(r_0)}
\newcommand{\pl}{{r_0}^{-\alpha}}

\newcommand{\pc}{p_{cov}}
\newcommand{\pcpcp}{p_{cov-c}}
\newcommand{\pcppp}{p_{cov-p}}
\newcommand{\popcp}{p_{out-c}}
\newcommand{\poppp}{p_{out-p}}
\newcommand{\pint}{P_{I}}
\newcommand{\rc}{R_c}

\newcommand{\ilos}{\mathbf{I}_{\text{LoS}}}
\newcommand{\iilos}{\mathbf{I}_{i,\text{LoS}}}
\newcommand{\inlos}{\mathbf{I}_{\text{NLoS}}}
\newcommand{\iinlos}{\mathbf{I}_{i,\text{NLoS}}}
\newcommand{\intlos}{\sum\limits_{i=1}^{\nlos} \mathbf{I}_{i,\text{LoS}}}
\newcommand{\intnlos}{\sum\limits_{i=1}^{\nnlos} \mathbf{I}_{i,\text{NLoS}}}

\newcommand{\prob}{\textbf{P}}
\newcommand{\SINR}{\text{SINR}}

\newcommand{\prcvd}{P_tG(\theta_0)h_0\plf}
\newcommand{\prcvdi}{P_tG(\theta_0)\plf}

\begin{document}

\title{V2X Downlink Coverage Analysis with a Realistic Urban Vehicular Model}
\author{\IEEEauthorblockN {Yae Jee~Cho,~Kaibin Huang*, and Chan-Byoung~Chae}\\
       \IEEEauthorblockA{School of Integrated Technology, Yonsei Institute of Convergence Technology, Yonsei University, Korea \\
       	*Department of Electrical \& Electronic Engineering, The University of Hong Kong, Hong Kong\\
      Email: \{yjenncho, cbchae\}@yonsei.ac.kr,~*huangkb@eee.hku.hk}}
\maketitle

\begin{abstract}
As the realization of vehicular communication such as vehicle-to-vehicle (V2V) or vehicle-to-infrastructure (V2I) is imperative for the autonomous driving cars, the understanding of realistic vehicle-to-everything (V2X) models is needed. While previous research has mostly targeted vehicular models in which vehicles are randomly distributed and the variable of carrier frequency was not considered, a more realistic analysis of the V2X model is proposed in this paper. We use a one-dimensional (1D) Poisson cluster process (PCP) to model a realistic scenario of vehicle distribution in a perpendicular cross line road urban area and compare the coverage results with the previous research that distributed vehicles randomly by Poisson Point Process (PPP). Moreover, we incorporate the effect of different carrier frequencies, mmWave and sub-6~GHz, to our analysis by altering the antenna radiation pattern accordingly. Results indicated that while the effect of clustering led to lower outage, using mmWave had even more significance in leading to lower outage. Moreover, line-of-sight (LoS) interference links are shown to be more dominant in lowering the outage than the non-line-of-sight (NLoS) links even though they are less in number. The analytical results give insight into designing and analyzing the urban V2X channels, and are verified by actual urban area three-dimensional (3D) ray-tracing simulation.\\ 
\end{abstract}
\begin{IEEEkeywords}
Urban vehicular models, V2X, downlink coverage, stochastic geometry, 3D ray-tracing.
\end{IEEEkeywords}


\section{Introduction}
The rapid advancements in the automobile industry are mainly directed to the autonomous driving cars which are expected to be present in the near future~\cite{niko2015auto}. The current autonomous cars in-development are expected to accompany a variety of sensors including radar, LiDAR (Light Detection and Ranging) and cameras which provide high quality but large sized data to aid self-driving technologies~\cite{Flem2012autotech}. The currently used DSRC (Dedicated Short-Range Communication)~\cite{kenn2011DSRC}, however, has only $1$km range and $2-6$ Mbps data-rate for communication which is not enough to support the at least 100Mbps of data-rate and high mobility required for the future connected cars with data heavy sensors. 
\begin{figure*}[t]
	\centering{\includegraphics[width=1\textwidth,keepaspectratio]
		{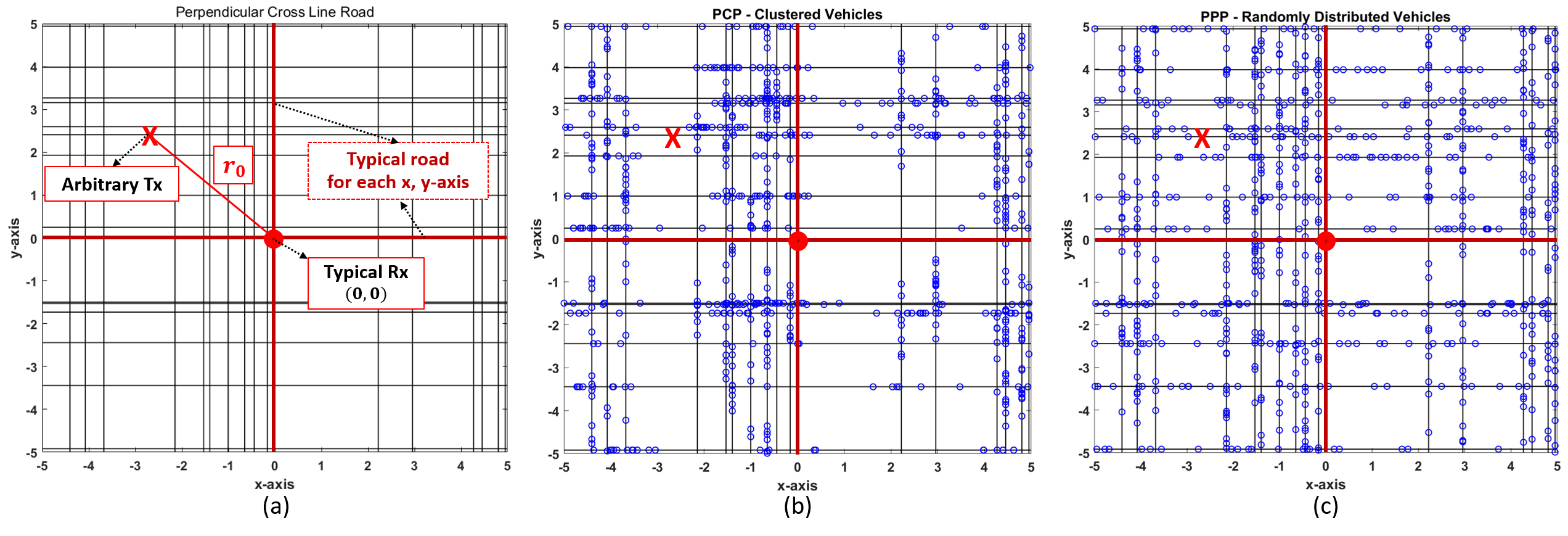}}
	\caption{An example of urban vehicular model with 731 and 745 vehicles for each PCP and PPP distributed vehicles in a $10\times10$ grid with unit $\SI{100}{\meter}$: (a) proposed system model of perpendicular cross line road with PPP modeled roads; (b) vehicles distributed by PCP; (c) vehicles distributed by PPP.} \label{f1:vehi_model} \vspace*{-3mm}
\end{figure*}

For such reasons, research on high mobility and data-rate communication for vehicle-to-everything (V2X) has been done in diverse perspectives. In \cite{choi2016mmVh}, Choi \etal{} used the measurements of DSRC and sensors to aid beam alignment of a roadside unit~(RSU) to randomly distributed vehicles, and results demonstrated decreased calculation overhead with negligible performance degradation. In \cite{stein2015stochastic}, Steinmetz \etal{} modeled a single and double pair of perpendicular lines using stochastic geometry to model randomly distributed vehicular communication channels near intersections. The outage probability is calculated for different distances amongst lines showing larger distance leads to lower outage probability. While such previous research give insight into vehicular communication modeling and implementation in urban areas, they target a certain vehicular condition in which vehicles are randomly distributed following the Poisson point process (PPP) or Erlang distribution. This gives much room for research on the vehicles distributed differently under different and more realistic conditions. In general, in actual urban areas there are many variables such as traffic lights, relative velocity, or mere traffic which causes the vehicles to be distributed in clusters rather than randomly~\cite{sham2011clus}. Moreover, while it is strongly expected that mmWave will be utilized for V2X for its large bandwidth resources~\cite{choi2016mmVh}, the comparison of its performance with the sub-6~GHz frequency range is also needed in an urban V2X scenario.  

In this paper, we present an urban V2X model using stochastic geometry~\cite{bai2014analysis,jeong2013interv,andrews2011tractable} and analyze the downlink coverage for a typical vehicle. Two independent homogeneous PPP for each x- and y-axis are used to model the roads. The vehicles are placed on each of the roads via a one-dimensional (1D) PPP or Poisson cluster process (PCP). RSUs are assumed to be on the streets along with the vehicles so can be considered as one of the nodes that are modeled as PPP or PCP for the vehicles. We assume Rayleigh fading channel with path-loss differently modeled for line-of-sight (LoS) and non-line-of-sight (NLoS) links, and consider different system settings for carrier frequency sub-6~GHz or mmWave. Based on such system, we analyze the coverage for a typical vehicle in terms of signal-to-interference-plus-noise ratio (SINR). Along with analytical results, actual simulation using three-dimensional (3D) ray-tracing is done to validate the results. 

The remainder of the paper is organized as follows. Section~II elaborates on the proposed urban vehicular model in terms of stochastic geometry. Section~III presents the system model of the V2X downlink channel along with the small-scale fading, path-loss, and antenna radiation pattern. Section~IV presents the coverage derivation for a typical vehicle based on the proposed system, and Section~V shows the results and analysis. Section~VI concludes the paper. 
\begin{table}[t]\label{notat}
	\centering \vspace{6mm}
	\caption{Notation}
	\renewcommand{\arraystretch}{1.3}
	\begin{tabular}{|p{1.5cm} |l|}
		\hline
		\bfseries{Symbol}& \bfseries{Definition} \\ 
		\hline 
		$N(\cdot)$ 	& Cardinality \\ \hline
		$\prob(\cdot)$ 	& Probability \\ \hline
		$\phi^P, \phi^C$   	& Poisson point process, Poisson cluster process\\ \hline
		$\lpar, \lr, \lv$   & Parent, road, and vehicle density    \\ \hline
		$p_i$ 		& Parent point of the $i^{\text{th}}$ cluster \\ \hline
		$c_i$		& Number of daughter points for the $i^{\text{th}}$ cluster\\ \hline
		$R, R_c$	& Range of entire grid and cluster\\ \hline
		$\sigma_c, \sigma_\theta, \sigma$	& Variance of cluster, antenna pattern, and noise\\ \hline
		$\nlos, \nnlos$				& Number of LoS roads and NLoS roads\\ \hline
		$v_0$				& $x$, $y$-coordinate of the typical vehicle\\ \hline
		$v_{i,j}$			& $x$, $y$-coordinate of the $j^{\text{th}}$ vehicle on the $i^{\text{th}}$ road\\ \hline
		$P_t$				& Transmit power\\ \hline
		$\pint$				& Interference probability\\
		\hline
	\end{tabular} 
	\vspace*{-3mm}
\end{table}
\section{Urban Vehicular Model}
\subsection{Perpendicular Cross Line Road}
To model a V2X scenario that reflects real urban areas, the geometry of the vehicular roads should be considered. For a strictly urban network, it is likely that the roads will be straight and perpendicular to each other, different from rural environments. Hence, a perpendicular cross line road model is proposed for representing the vehicular roads in urban environments. The model appears as in Fig.~\ref{f1:vehi_model}(a). For a grid of fixed size of $R \times R \; (R = 10$ in Fig.~\ref{f1:vehi_model}, unit $\SI{100}{\meter}$), we first model the roads as two independent and identically distributed (i.i.d.) homogeneous PPP $\xppp$,$\yppp$, for each $x-$ and $y-$axis where,
\begin{align}
\xppp,\yppp \sim \prob(N(R) = n_r) = \frac{e^{R\lr}{R\lr}^{n_r}}{n_r!}.
\label{ppproad}
\end{align} 
\vspace{-2mm} \\
$n_r$ is the number of roads for a road of length $R$ and $\lr$ is the unit number of roads per length, i.e., density for the number of roads. $\lr$ is set constant since the number of roads do not change in general. We fix a typical road at the center of the grid for each $x-$ and $y-$axis which are independent from $\xppp$ and $\yppp$ (See Fig.~\ref{f1:vehi_model}(a)). The intersection of these typical roads are the coordinates of the typical vehicle, i.e., $v_0 = (0,0)$. The typical roads are considered as the LoS roads where the interference links arriving at the typical vehicle from the vehicles on these roads are LoS-dominant. The other roads are all considered as NLoS-dominant, which is applicable to urban areas due to the high buildings and frequent obstacles. Hence in our vehicular channel model $\nlos = 2, \nnlos=n_r-\nlos$. \footnote{Note that the $\xppp$ and $\yppp$ processes used to model the horizontal and vertical roads are symmetric in the perspective of the typical vehicle. Therefore, when calculating the interference links for the typical vehicle we can simply get the sum of all the roads regardless of the horizontal and vertical axis.} 

\subsection{Vehicular Node Distribution}
While it is common to model the vehicles as randomly distributed, in actual situations vehicles are rather grouped and clustered due to the relative velocity and intersections. Hence we model the vehicles into two cases: randomly and clustered as in Fig.~\ref{f1:vehi_model}(b) and (c) and compare the results accordingly. In the proposed V2X model, the base stations (infrastructures) are RSUs which are placed on the streets along with the vehicles so can be considered as one of the nodes that are modeled as PPP or PCP for the vehicles. 

First, to model the case of the clustered vehicles for each road, the vehicles are modeled as a 1D Neyman-Scott homogeneous PCP, $\pcp$, where the parent points follow the stationary Poisson process of $\pppp=\left\{ p_1,p_2,...,p_i,...\right\}$ with intensity $\lpar$ similarly defined as~(\ref{ppproad}). Each parent point $p_i$ represents the center of the $i^{\text{th}}$ cluster which has $c_{i}$ daughter points where $c_{i}$ is Poison distributed with mean $\cbar$, i.e., $c_{i}\sim \text{Pois}(\cbar)$. Since vehicles are tended to be distributed more densely at the center of the intersection, or a certain cluster, we employ the Thomas cluster process where each daughter point $x$ is distributed around the parent point $p_i$ as a symmetric normal distribution with variance $\csig^2$, i.e., \\ \vspace{-4mm}
\begin{align}
f(d_i)=\frac{1}{2\pi\sigma_c^2}\texp\left( -\frac{d_i^2}{2\csig^2}\right)
\end{align} \vspace{-1mm} \\
where $d_i$ is the distance between parent point $p_i$ and one of its daughter point $x~(d_i=||p_i-x||)$, and $\csig^2$ is the variance of the daughter points\apo~location from their parent point.

Secondly, to model the case of the randomly distributed vehicles, 1D PPP vehicle model $\vppp$ is used as in Fig.~\ref{f1:vehi_model}(c) also similarly defined as (\ref{ppproad}) with density $\lv$. Note that the Poisson mean for $\pcp$ and $\vppp$ are $R\lpar$, $R\lpar\cbar$ respectively which makes the average number of vehicles identical between the clustered and random vehicular model. 
\vspace{2mm}
\begin{figure}[t]
	\centering{\includegraphics[width=0.9\columnwidth,keepaspectratio]
		{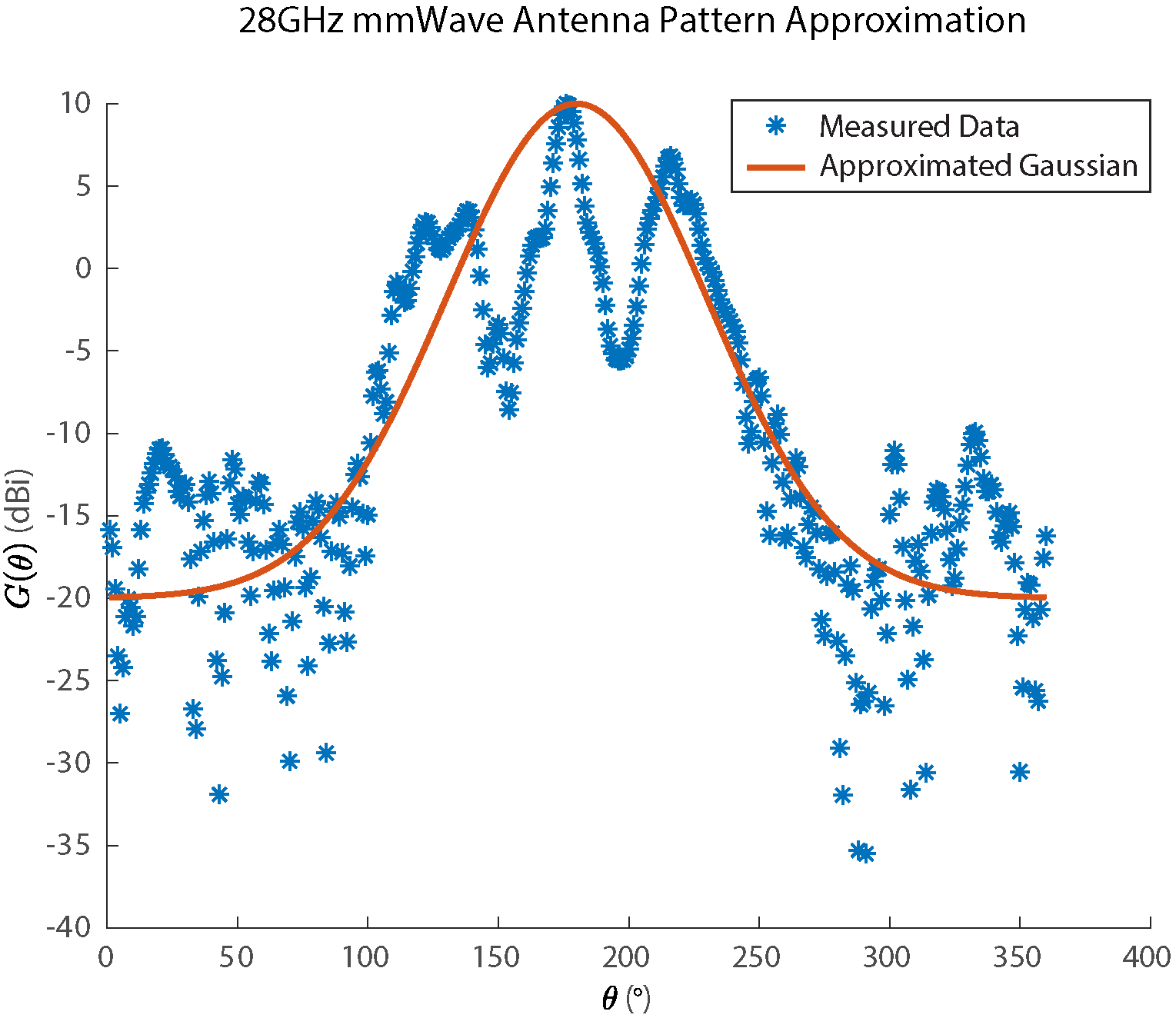}}
	\caption{Antenna pattern approximation by Gaussian for an actual 28~GHz lens antenna\apo s radiation pattern. \vspace{4mm}} \label{f2:ant_pat}
\end{figure}
\section{System Model}
In this section, we present the general system model for the channel of our vehicular model including the small-scale fading, path-loss, and antenna radiation pattern which are adjusted differently to the cases when the carrier frequency is either mmWave or sub-6~GHz. Note that although in future 5G networks multi-connectivity is expected, and hence it is most likely that sub-6~GHz and mmWave will both be used in a system, we assume each case separately in this paper.

 \begin{figure*}[!b]
	\begin{align*}
	\texp\left(-\pint\lpar \int\displaylimits_{-R}^{R}\left[1-M\left(\int\displaylimits_{-\rc}^{\rc}\int\displaylimits_{0}^{2\pi}f(||y||)(1+sP_tG(\theta_x)\ell(||x+y||))^{-1}p_{AoA}(\theta_x)\,\mathrm{d}\theta_x\,\mathrm{d}y\right)\right]\mathrm{d}x\right). \tag{16}
	\label{SINR_PCPlos}
	\end{align*}
\end{figure*}

\subsection{Downlink Received Power}
We consider the downlink scenario for the typical vehicle $v_0$ located at the center of the grid where the other nodes (which can be other transmitting vehicles or RSUs) are potential interference signals. The typical vehicle is assumed to have a LoS dominant link of length $r_0$. Then, the received power for the typical vehicle can be presented as, 
\begin{align}
P_\text{rcvd} = \prcvd
\end{align}
where $P_t, G(\theta_0), h_0, \plf$ correspond to the transmit power, antenna gain at the main beam direction $\theta_0$, small-scale fading, and path-loss respectively. The transmit power is assumed to be identical for all nodes as $P_t = 1$ for convenience, and the antenna gain is elaborated in detail in the Section~\ref{ant_sec}. For the small-scale fading factor we assume Rayleigh fading for all scenarios and $\plf$ is the path-loss for LoS environment defined by the \textit{general distance based path-loss law} as $\plf=\pl$ where the path-loss exponent is $\alpha=2$. Log-normal shadowing is assumed to be negligible in our scenario since the value is small compared to the interference and noise power we will evaluate for the SINR. 

If $\phi$ can represent either $\pcp$ or $\vppp$, the interference signal coming from a $j^{\text{th}}$ node on the $i^{\text{th}}$ LoS road, i.e., $(i,j)^\text{th}$ LoS link, is as follows:
\begin{align}
\ilos=\intlos=\sum\limits_{i=1}^{\nlos}\sum_{j\in \phi} P_tG(\theta_{i,j})h_{i,j}\ell(r_{i,j})
\label{intlos}
\end{align}\vspace{-2mm} 

\noindent{where $r_{i,j}=||v_{i,j}-v_0||$, i.e., the distance between the typical vehicle and the interference node, and $\theta_{i,j}$ is the angle-of-arrival for the $(i,j)^\text{th}$ LoS link. For the interference signal from a $j^{\text{th}}$ node on the $i^{\text{th}}$ NLoS road, it is most likely that there are obstacles which cause blockage loss to be much more dominant than the distance-based propagation loss~\cite{baccelli2015correlated}; this becomes more likely in the mmWave domain. Hence, for the interference signal link from the NLoS road, we consider a \textit{blockage path-loss model}, and the interference signal becomes,} 
\begin{align}
\inlos=\intnlos=\sum\limits_{i=1}^{\nnlos}\sum_{j\in \phi} P_tG(\theta_{i,j})h_{i,j}L^{K_{i,j}}.
\label{intnlos}
\end{align}\vspace{-2mm}

\noindent{$L\in(0,1)$ is the penetration loss for the corresponding environment, and $K_{i,j}$ is the number of buildings that the $(i,j)^\text{th}$ NLoS link crosses. The value of $L$ will be high for the mmWave case and low for the sub-6~GHz. For an arbitrary NLoS link of length $r_{i,j}$ and angle-of-arrival $\theta_{i,j}$, $K_{i,j}-1$ is Poisson distributed as,  $K_{i,j}-1\sim\text{Pois}(R\lpar\cbar r_{i,j}(|\cos\theta_{i,j}|+|\sin\theta_{i,j}|))$~\cite{baccelli2015correlated}. The value of penetration loss $L$ depends on the carrier frequency and building materials which value we will use from previous measurement studies on penetration loss.
}

\subsection{Antenna Gain} \label{ant_sec}
For an arbitrary angle $\theta$, we define the antenna gain, $G(\theta)$, differently for mmWave and sub-6~GHz carrier frequency. The antenna radiation pattern when using mmWave is much narrower with higher directivity and gain. While it is very common to use nearly omnidirectional antennas for the sub-6~GHz which antenna gain is close to uniform for all directions, such feature cannot be implemented in mmWave. In Fig.~\ref{f2:ant_pat}, we present the actual antenna radiation pattern in vertical polarization of a 28~GHz mmWave lens antenna~\cite{yjc_mmWavelens,kwon2016rf}. We approximate this actual measured pattern to a normal distribution probability density function (pdf)~\cite{zhang2017gau} as $N\sim(\mu_\theta,{\sigma_\theta}^2)$ where $\mu_\theta=180^\circ$, $\sigma_\theta=50^\circ$.
For the sub-6~GHz carrier frequency, we derive the antenna pattern equation as an optimal omni-directional antenna pattern, i.e., uniform gain for all angles. The total gain is identical for both mmWave and sub-6~GHz cases.

\section{Downlink Coverage Analysis}
Based on the vehicle distribution and system model mentioned above, we derive the coverage probability of the typical vehicle for different cases. The coverage probability, $\pc$, for a predefined threshold $T$ can be defined as, 
\vspace{-3mm}

\begin{align}
&\pc = \prob[\SINR > T]\\[1.7ex]
&=\prob\left(\frac{P_\text{rcvd}}{\sigma^2+\ilos+\inlos}>T\right)\\[1.7ex]
&=\prob\left(\frac{\prcvd}{\sigma^2+\ilos+\inlos}>T\right)\\[1.7ex]
&=\E_{\ilos,\inlos}\left[\text{Pr}\left(h>\frac{T(\sigma^2+\ilos+\inlos)}{\prcvdi}\right)\right] \label{SINR_def}\\[1.7ex]
&=\texp(-s\sigma^2)\E_{\ilos,\inlos}\left[\texp(-s(\ilos+\inlos)\right)]\label{SINR_def_1}\\[1.7ex]
&=\texp(-s\sigma^2)\prod_{i=1}^{\nlos} \underbrace{\mathcal{L}_{\iilos}(s)}_\textrm{(i)}\prod_{i=1}^{\nnlos} \underbrace{\mathcal{L}_{\iinlos}(s)}_\textrm{(ii)}. \label{SINR_def_2}
\end{align}


\noindent Eq. (\ref{SINR_def_1}) is derived by the properties of the Rayleigh small-scale fading, $h_0, h_{i,j}\sim\texp(1)$, and defining $s=T/(\prcvdi)$. Finally, due to the independence of the variable $\ilos$ and $\inlos$, we can use the property of the Laplace transform to derive (\ref{SINR_def_1}) as~(\ref{SINR_def_2}).

For both the randomly placed and clustered vehicles, we can derive the LoS interference, i.e. (i) from (\ref{SINR_def_2}) as, \vspace{2mm}
\begin{align}
&{\mathcal{L}_{\iilos}(s)}=\E_{\iilos}[\texp(-s\iilos)] \label{eq1}\\[1.5ex]
&=\E_{\iilos}\left[\texp(-s\sum_{j\in \phi} P_tG(\theta_{i,j})h_{i,j}\ell(r_{i,j}))\right] \label{eq2}\\[1.5ex]
&=\E_{\iilos}\left[\prod_{j\in \phi}\E_h(\texp(-sP_tG(\theta_{i,j})h_{i,j}\ell(r_{i,j})))\right] \label{eq4}\\[1.5ex]
&=\E_{\iilos}\left[\prod_{j\in \phi}\frac{1}{1+sP_tG(\theta_{i,j})\ell(r_{i,j})}\right]. \label{eq5}
\end{align}
\begin{figure*}[t]
	\centering{\includegraphics[width=1\textwidth,keepaspectratio]
		{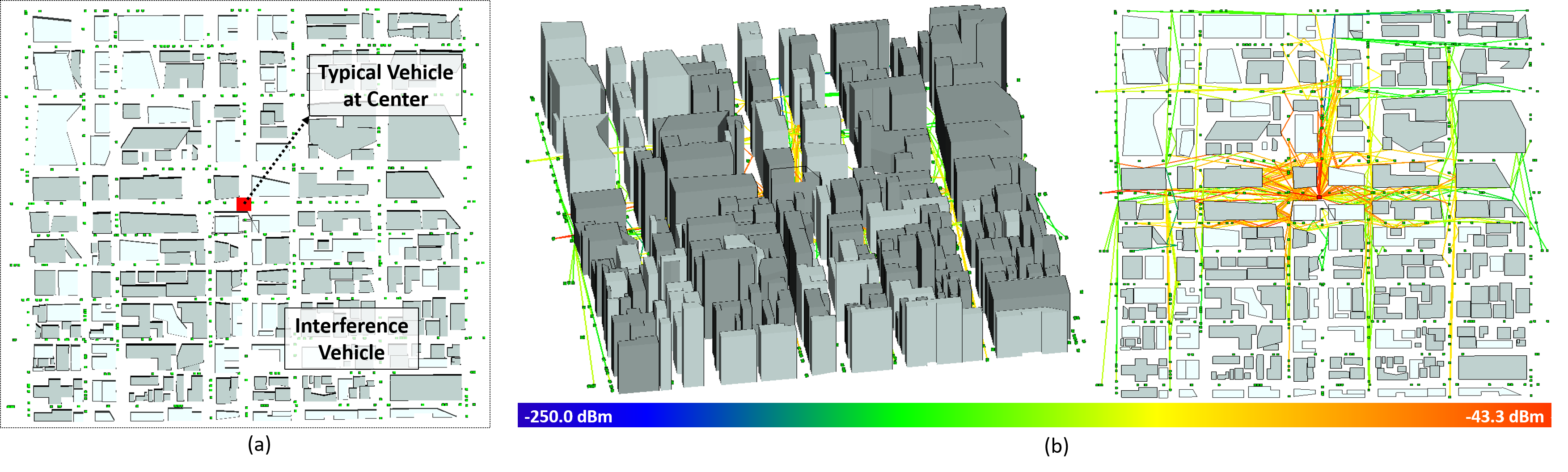}}
	\caption{Urban vehicular model 3D ray-tracing simulation environment: (a) top view; (b) side view and top view with example of ray-tracing.} \label{fig0:ray_3D}
\end{figure*}
\noindent Since the form of (\ref{eq5}) can be applied to the probability generating functional~\footnote{For an arbitrary function $v(\cdot)$, the probability generating functional is defined as $\E\left(\prod_{j\in \phi}v(j)\right)$.}~\cite{chiu2013stoc}. for the clustered case ($\phi \rightarrow \pcp$), (\ref{eq5}) is derived as (\ref{SINR_PCPlos}) where $M(z)=\texp(-\cbar(1-x))$ is the moment generating function for the Thomas cluster process~\cite{henggi2009clust}, and $p_{AoA}(\theta_x)$ is the pdf of the angle-of-arrival defined as $\theta_x\sim~U(0,2\pi)$~\cite{rapp2014mmWch}. Similarly, using the probability generating functional for the PPP, for the randomly distributed vehicles~($\phi \rightarrow \vppp$) we can derive (i) from Eq.~(\ref{eq5}) as (\ref{SINR_PPPlos}).
\vspace{2mm}
\begin{flalign*}
\texp\left(\!-\pint\lv\int\displaylimits_{-R}^{R}\!\int\displaylimits_{0}^{2\pi}\!\frac{p_{AoA}(\theta_x)}{1+(sP_tG(\theta_x)\ell(||x||))^{-1}}\mathrm{d}\theta_x\mathrm{d}x\right). \tag{17}
\label{SINR_PPPlos}
\end{flalign*}

\noindent The NLoS interference (ii) from (\ref{SINR_def_2}) for each clustered and random case is in the same form as (\ref{SINR_PCPlos}) and (\ref{SINR_PPPlos}) where $\ell(\cdot)$ is replaced by the blockage path-loss model $L^K_{i,j}$. By plugging the corresponding equations into (\ref{SINR_def_2}), we can get the coverage probability for each of the randomly distributed and clustered vehicles, i.e., $\pcppp,~\pcpcp$. Moreover we modulate the cases for each sub-6~GHz and mmWave by setting $G(\theta)$ and $L$ differently as mentioned in Section~III. \label{cov_ana}

\section{Result Analysis with 3D Ray-Tracing Simulation}
\subsection{Simulation and Parameters}
We utilize the derived analytical expressions of the coverage probability to get the outage probability for each randomly distributed case and clustered case as $\poppp=1-\pcppp$ and $\popcp=1-\pcpcp$, and analyzed the effect of clustering, carrier frequency, and other parameters, i.e., $\cbar,~r_o,$ and $T$ on the outage probability. Note that our analytical expression for the coverage probability, (\ref{SINR_def_2}), can be separated into three parts in the perspective of LoS and NLoS: when there is only LoS roads ($=\texp(-s\sigma^2)\prod_{i=1}^{\nlos}\mathcal{L}_{\iilos}(s)$), only NLoS roads ($=\texp(-s\sigma^2)\prod_{i=1}^{\nnlos}\mathcal{L}_{\iinlos}(s)$), both LoS and NLoS roads ($=\texp(-s\sigma^2)\prod_{i=1}^{\nlos}\mathcal{L}_{\iilos}(s)\prod_{i=1}^{\nnlos}\mathcal{L}_{\iinlos}(s)$).\footnote{Since the urban vehicular model has high obstacles and confined roads, we consider the two typical roads as the LoS roads ($\nlos=2$), and the rest of the roads as the NLoS roads($\nnlos=n_r-\nlos$).} In order to see the effect of the blockages in an urban vehicular environment, we consider all three cases separately in our analysis. 

\begin{table}[t]\vspace{5mm}
	\begin{center} 
		\caption{Simulation Parameters}
		\renewcommand{\arraystretch}{1.4}
		\begin{tabular}{|p{5cm}|c|}
			\hline
			\bfseries{Parameter} 							& \bfseries{Value} \\ 
			\hline 
			Range of entire grid and cluster, $R, R_c$			& ${5, 1}$ \\ \hline
			LoS path-loss exponent, $\alpha$	& 2\\ \hline
			Parent density, $\lpar$	& 0.5 \\ \hline
			Mean number of nodes in cluster, $\cbar$ & 5 \\ \hline
			Variance of cluster, $\sigma_c$ & 0.5 \\ \hline
			Transmit power, $P_t$	&	43~dBm \\ \hline
			Probability of interference, $P_I$	&	0.3 \\ \hline
			\multirow{ 2}{*}{Penetration loss, $L$}	& -40~dB~(\text{for mmWave}) \\ &-30~dB~(\text{for sub-6~GHz})\\ \hline
			Number of LoS roads, $\nlos$	&	2 \\ \hline
			Number of NLoS roads, $\nnlos$	&	$n_r-\nlos$ \\ \hline
			Noise variance, $\sigma^2$	&  -104.5~dBm\\
			\hline \end{tabular} 
	\end{center}
	\vspace*{-6mm}
\end{table}
\begin{figure}[t]
	\centering{\includegraphics[width=0.81\columnwidth,keepaspectratio]
		{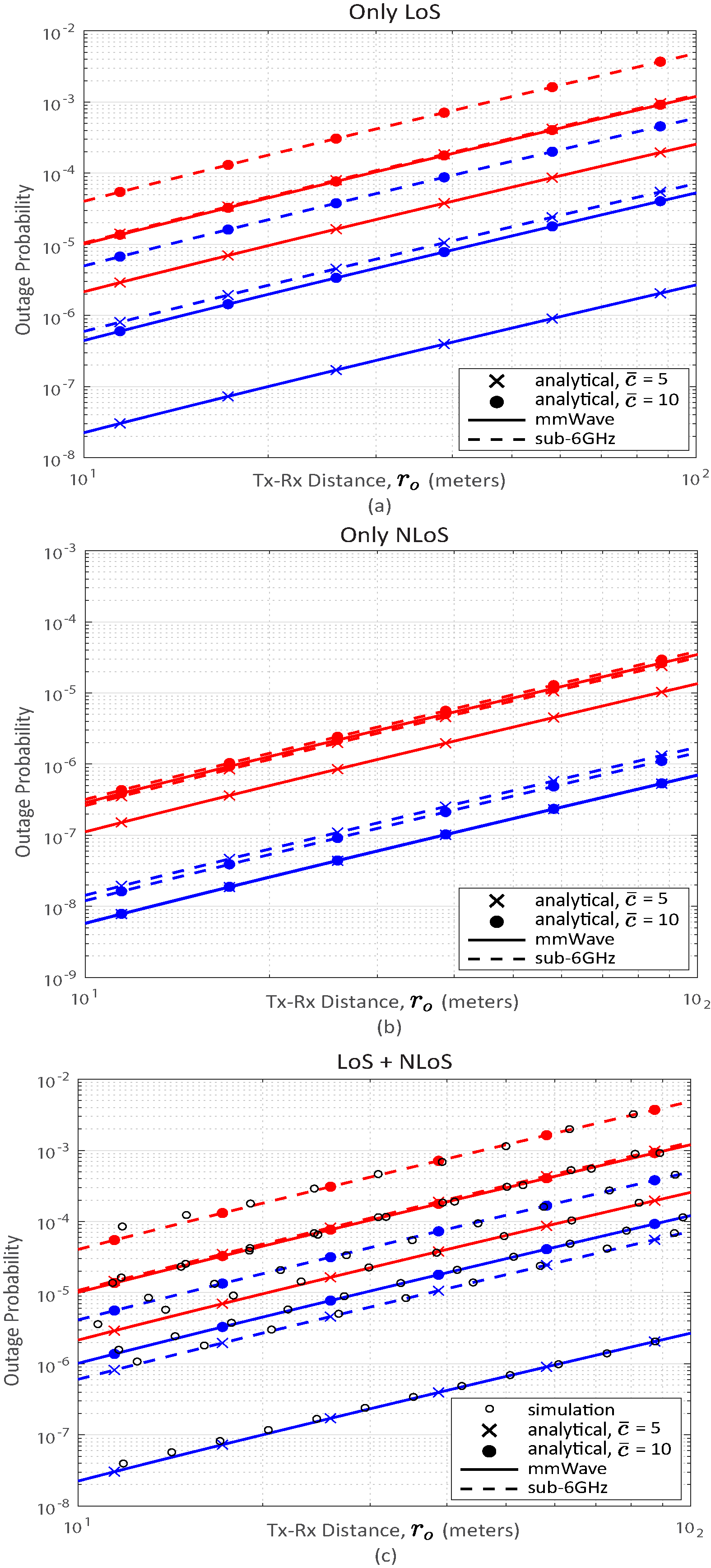}}
	\caption{\small{Outage probability as a function of $r_o$ for different $\cbar$ and carrier frequencies (Blue: PCP distributed vehicles, Red: PPP distributed vehicles): (a) only LoS; (b) only NLoS; (c) Both LoS and NLoS ($T=-10~\text{dBm},~ \sigma_c=0.8$).}} \label{fig1:cbar}
	\vspace{-0.5mm}
\end{figure}
\begin{figure}[t]
	\centering{\includegraphics[width=0.9\columnwidth,keepaspectratio]
		{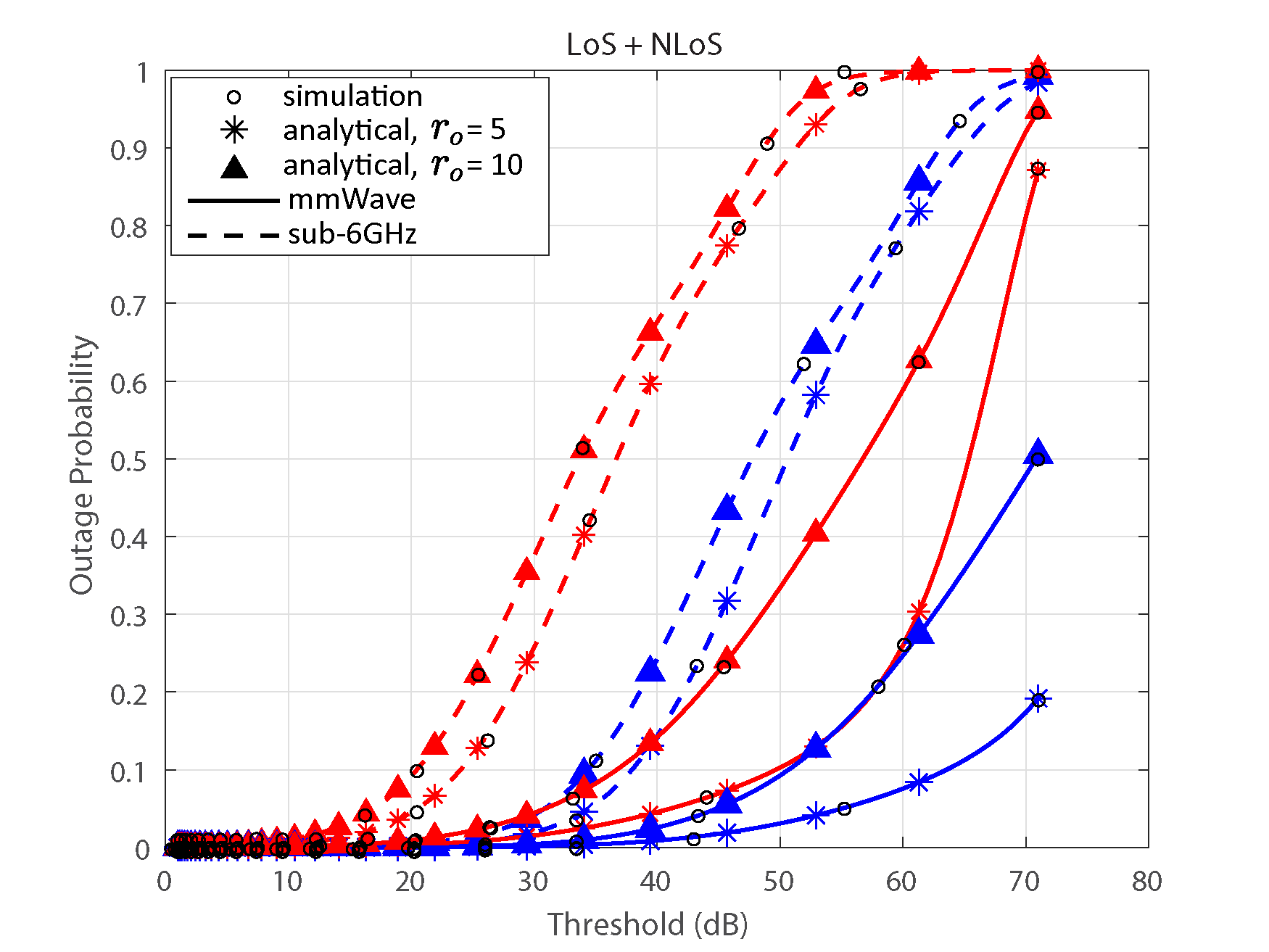}}
	\caption{\small{Outage probability as a function of threshold $T$ for different $r_o$ and carrier frequencies (Blue: PCP distributed vehicles, Red: PPP distributed vehicles; $\cbar=5,~ \sigma_c=0.8$).}} 
	\label{fig2:thre}
\end{figure}
To validate the analytical coverage probability, we use Wireless InSite 3D ray-tracing tool to simulate the urban vehicular environment used in the analytical model and compare the results from the simulation and analytical derivation. As in Fig.~\ref{fig0:ray_3D}, we modeled an actual region from Ottawa, Canada ($\SI{1}{\kilo\meter}\times\SI{1}{\kilo\meter}$). For each instance of simulation trial, the roads and obstacles are fixed, but the vehicles placed on the roads are generated by either $\pcp$ or $\vppp$. All vehicles except the typical vehicle placed at the center of the area can be potential interference vehicles with probability $P_I$. The building obstacles are built of concrete material. For each type of simulation, for example PPP distributed vehicles with sub-6~GHz frequency, we generated 10 different instances of trials and averaged the outage probability results. In the simulation, we only considered the both LoS and NLoS roads case without loss of generality. As identical to the analytical evaluation, the antenna radiation pattern is changed accordingly in the simulation when the carrier frequency is changed to either mmWave or sub-6~GHz: 28~GHz lens antenna for mmWave and omnidirectional antenna for sub-6~GHz. The direction in which the mmWave beam for each vehicle is aligned is chosen randomly. The simulation parameters are listed in~Table.II. Note that the simulation parameters were set to identical values with those used for the analytical expression for the coverage probability.

\subsection{Results and Analysis}
In Fig.~\ref{fig1:cbar} for all of the only LoS, only NLos, and LoS+NLoS cases we can see that the clustering of the vehicles, i.e., vehicles following $\pcp$, give less interference to the typical vehicle resulting in a lower outage than vehicles following $\vppp$. Also, the outage decreases as $\cbar$ is decreased since less clustering and number of interference occurs. This implies that the clustering can give an effect of decreasing the number of interference vehicles that can be closer to the typical vehicle thus decreasing the outage. Moreover, for both cases of the vehicles following $\pcp$ and $\vppp$, the outage is lower for the case when mmWave is used as the carrier frequency. This indicates that for V2X downlink scenario in urban areas, the blockages have a significant effect because interference links are blocked in mmWave. 

Now if we compare the cases of (a),(b), and (c) of Fig.~\ref{fig1:cbar} we get an interesting result that for both PPP and PCP cases and sub-6~GHz and mmWave cases, the only LoS case has a higher outage probability even if the number of total interference vehicles is very likely to be smaller than the only NloS case. This indicates that for the proposed urban vehicular environment, the LoS interference signals have a greater effect to the communication link rather than the NLoS link. This is also seen in (c) since the trend of the outage probability is very much similar to the only LoS case than the only NLoS case. Lastly, we can see from (c) that the simulation data from 3D ray-tracing fits well with the analytical results. In Fig.~\ref{fig2:thre}, for the increasing thresholds, the outage probability increases for all cases. Similar to Fig.~\ref{fig1:cbar} the outage probability is lower for mmWave systems, and the vehicles following $\pcp$ leads to lower outage. 

%
\section{Conclusion}
In this paper, we presented an urban vehicular model by using the PPP and PCP. By considering realistic urban V2X scenarios in which vehicles have group mobility and are clustered, we analyzed the outage probability for a typical vehicle. The effect of clustering and using either mmWave or sub-6~GHz in the V2X model is analyzed. Results showed that the clustering of vehicles lowers the outage by decreasing the number of interference vehicles that can be closer to the typical vehicle. Moreover, using mmWave as the carrier frequency yielded lower outage due to more NLoS interference links being blocked by obstacles and less interference power for LoS interference links due to the mmWave narrow beamwidth. Another significant result was derived that although NLoS roads are more dominant in number, the LoS interference link had more significance to the outage giving an insight in to V2X channels. Lastly, as outage increases with increasing threshold, we also demonstrated that the effect of carrier frequency more significantly impacts the outage than the clustering effect. The analytical results were compared with the 3D ray-tracing simulation results, and were verified to match well. 

\section*{Acknowledgment}
\small This work was supported by IITP grant funded by the Korea government~(MSIT) (2018-0-00208, High Accurate Positioning Enabled MIMO Transmission and Network Technologies for Next 5G-V2X~(vehicle-to-everything) Services).

\bibliography{networkRefs}

\begin{thebibliography}{10}
\providecommand{\url}[1]{#1}
\csname url@samestyle\endcsname
\providecommand{\newblock}{\relax}
\providecommand{\bibinfo}[2]{#2}
\providecommand{\BIBentrySTDinterwordspacing}{\spaceskip=0pt\relax}
\providecommand{\BIBentryALTinterwordstretchfactor}{4}
\providecommand{\BIBentryALTinterwordspacing}{\spaceskip=\fontdimen2\font plus
\BIBentryALTinterwordstretchfactor\fontdimen3\font minus
  \fontdimen4\font\relax}
\providecommand{\BIBforeignlanguage}[2]{{%
\expandafter\ifx\csname l@#1\endcsname\relax
\typeout{** WARNING: IEEEtran.bst: No hyphenation pattern has been}%
\typeout{** loaded for the language `#1'. Using the pattern for}%
\typeout{** the default language instead.}%
\else
\language=\csname l@#1\endcsname
\fi
#2}}
\providecommand{\BIBdecl}{\relax}
\BIBdecl

\bibitem{niko2015auto}
M.~Nikowitz, ``Fully autonomous vehicles: Visions of the future or still
  reality?'' epubli GmbH, 2015, available:
  \url{https://resources.ext.nokia.com/asset/200176}.

\bibitem{Flem2012autotech}
B.~Fleming, ``New automotive electronics technologies,'' \emph{{IEEE} Veh.
  Technol. Mag.}, vol.~7, no.~4, pp. 4--12, Dec 2012.

\bibitem{kenn2011DSRC}
J.~B. Kenney, ``Dedicated short-range communications {(DSRC)} standards in the
  united states,'' \emph{Proc. {IEEE}}, vol.~99, no.~7, pp. 1162--1182, June
  2011.

\bibitem{choi2016mmVh}
J.~Choi, V.~Va, N.~Gonz{\'a}lez-Prelcic, R.~Daniels, C.~R. Bhat, and R.~W.~H.
  Jr., ``Millimeter-wave vehicular communication to support massive automotive
  sensing,'' \emph{{IEEE} Commun. Mag.}, vol.~54, no.~12, pp. 160--167, Dec
  2016.

\bibitem{stein2015stochastic}
E.~Steinmetz, M.~Wildemeersch, T.~Quek, and H.~Wymeersch, ``A stochastic
  geometry model for vehicular communication near intersections,'' in
  \emph{Proc. {IEEE} {GLOBECOM} Workshops (GC Wkshps)}, Dec 2015, pp. 1--6.

\bibitem{sham2011clus}
D.~Shamoto, A.~Tomoeda, R.~Nishi, and K.~Nishinari, ``Car-following model with
  relative-velocity effect and its experimental verification,'' \emph{Phys.
  Rev. E}, vol.~83, p. 046105, April 2011.

\bibitem{bai2014analysis}
T.~Bai, R.~Vaze, and {R. W. Heath Jr.}, ``Analysis of blockage effects on urban
  cellular networks,'' \emph{{IEEE} Trans. Wireless Commun.}, vol.~13, no.~9,
  pp. 5070--5083, June 2014.

\bibitem{jeong2013interv}
Y.~Jeong, J.~W. Chong, H.~Shin, and M.~Z. Win, ``Intervehicle communication:
  Cox-fox modeling,'' \emph{{IEEE} J. Sel. Areas Commun.}, vol.~31, no.~9, pp.
  418--433, July 2013.

\bibitem{andrews2011tractable}
J.~G. Andrews, F.~Baccelli, and R.~K. Ganti, ``A tractable approach to coverage
  and rate in cellular networks,'' \emph{{IEEE} Trans. Commun.}, vol.~59,
  no.~11, pp. 3122--3134, Oct. 2011.

\bibitem{baccelli2015correlated}
F.~Baccelli and X.~Zhang, ``A correlated shadowing model for urban wireless
  networks,'' in \emph{Proc. {IEEE} Conf. on Comput. Commun. ({INFOCOM})},
  April 2015, pp. 801--809.

\bibitem{yjc_mmWavelens}
Y.~J. Cho, G.~Y. Suk, B.~Kim, D.~K. Kim, and C.-B. Chae, ``{RF} lens-embedded
  antenna array for mmwave {MIMO}: Design and performance,'' \emph{{IEEE}
  Commun. Mag.}, to be published.

\bibitem{kwon2016rf}
T.~Kwon, Y.-G. Lim, B.-W. Min, and C.-B. Chae, ``{RF} lens-embedded massive
  {MIMO} systems: Fabrication issues and codebook design,'' \emph{{IEEE} Trans.
  Microw. Theory Tech.}, vol.~64, no.~7, pp. 2256--2271, Jun 2016.

\bibitem{zhang2017gau}
X.~Zhang and C.~D. Sarris, ``A {Gaussian} beam approximation approach for
  embedding antennas into vector parabolic equation-based wireless channel
  propagation models,'' \emph{{IEEE} Trans. Antennas Propag.}, vol.~65, no.~3,
  pp. 1301--1310, March 2017.

\bibitem{chiu2013stoc}
S.~N. Chiu, D.~Stoyan, W.~S. Kendall, and J.~Mecke, \emph{Stochastic Geometry
  and its Applications}.\hskip 1em plus 0.5em minus 0.4em\relax Hoboken, NJ,
  USA: Wiley, 2013.

\bibitem{henggi2009clust}
G.~Radha.~K and H.~Martin, ``Interference and outage in clustered wireless
  \textit{ad hoc} networks,'' \emph{{IEEE} Trans. Inf. Theory}, vol.~55, no.~9,
  pp. 4067--4086, Aug. 2009.

\bibitem{rapp2014mmWch}
A.~Mustafa~R., L.~Yuanpeng, S.~Mathew~K., S.~Sun, S.~Rangan, T.~S. Rappaport,
  and E.~Erkip, ``Millimeter wave channel modeling and cellular capacity
  evaluation,'' \emph{{IEEE} J. Sel. Areas Commun.}, vol.~32, no.~6, pp.
  1164--1179, June 2014.

\end{thebibliography}
\bibliographystyle{IEEEtran} 

\end{document}